\documentclass[twocolumn]{revtex4}
\pdfoutput=1
\usepackage{amssymb}
\usepackage{amsmath}
\usepackage{graphicx}
\usepackage[english]{babel}
\begin{document}
\title{Electronic and optical properties of carbon nanodisks and nanocones}
\author{P. Ulloa}
\email{pablo.ulloa@usm.cl}
\author{M. Pacheco}
\affiliation{Departamento de F\'isica, Universidad T\'ecnica Federico Santa Mar\'ia, Casilla 110-V, Valpara\'iso, Chile}
\author{L. E. Oliveira}
\affiliation{Instituto de F\'isica, Universidade Estadual de Campinas-UNICAMP, Campinas-SP, 13083-859, Brazil}
\author{A. Latg\'e}
\affiliation{Instituto de F\'isica, Universidade Federal Fluminense, 24210-340, Niter\'oi-RJ, Brazil}
\date{\today}
\begin{abstract}
A theoretical study of the electronic properties of nanodisks and nanocones is presented within the framework of a tight-binding scheme. The electronic densities of states and absorption coefficients are calculated for such structures with different sizes and topologies. A discrete position approximation is used to describe the electronic states taking into account the effect of the overlap integral to first order. For small finite systems, both total and local densities of states depend sensitively on the number of atoms and characteristic geometry of the structures. Results for the local densities of charge reveal a finite charge distribution around some atoms at the apices and borders of the cone structures. For structures with more than 5000 atoms, the contribution to the total density of states near the Fermi level essentially comes from states localized at the edges. For other energies the average density of states  exhibits similar features to the case of a graphene lattice. Results for the  absorption spectra of nanocones show a peculiar dependence on the photon  polarization in the infrared range for all investigated structures.
\end{abstract}
\maketitle
\section{Introduction}

Carbon nanocones (CNCs) have been observed by microscopy techniques\cite{Sattler1995,Garberg2008}, showing an ordered atomic structure.
Large progress has been made on synthesis, characterization and manipulation of CNCs and carbon nanodisks (CNDs) \cite{Naess2009,Lin2007,Krishnan1997,Zhang2009,delCampo2013,Ritter2009nature}. From the technological point of view, applications such as microscopy probes and electron-emitter devices may be envisaged by considering the electric current established through the cone apex when electric fields are applied\cite{Houdellier2012}.

There are different theoretical schemes to describe the electronic properties of cone-like structures.
Models based on the  Dirac equation\cite{Lammert2004,Sitenko2008} give a convenient insight of properties in the long wavelength limit. However, for  finite-size graphenes, the longest stationary wavelength occurs in the border and a correct description of the states near the Fermi level is given in terms of edge states\cite{Wimmer2010,Nakada1996}.
Ab initio models \cite{Heiberg2008,Kobayashi2000,Ming2012} are able to predict detailed features, but they are restricted to structures composed of a few hundred atoms due to their considerable computational costs.
Calculations based on a single $\pi$ orbital are able to describe the relevant electronic properties\cite{Chen2010,Tamura1994}. In that spirit, we calculate the electronic structure and optical spectra of CNDs and CNCs within a tight binding approach.  CNC structured systems generated by pentagonal and heptagonal defects were previously studied using a Green function recursive method \cite{Tamura1994,Tamura1997}. It was shown that, for cones generated with an odd disclination number $n_w$, it is not possible to define A and B graphene sublattices. In this case, therefore, the electron-hole symmetry is broken.

\begin{figure}
\includegraphics[width=\linewidth]{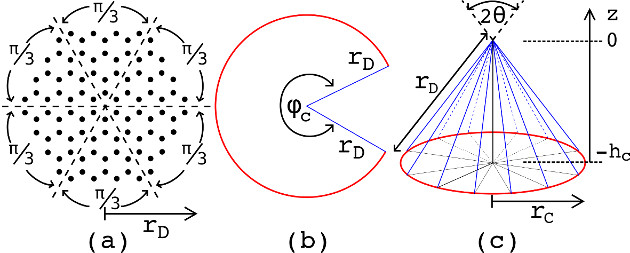}
\caption{\label{fig:pictorial-view}(Color online)
Pictorial view of (a) a carbon disk composed of six wedge-section of $\pi/3$ central angle, then (b) the remotion of a $n_w=\{1,\dots,5\}$ wedge-sections from the disk, and (c) by folding it is constructed a cone. Geometrical elements: generatrix $r_D$, height $h_c$, base radius $r_c$,
and apex opening angle $2\theta$, where $\sin\theta = 1-n_w/6$.}
\end{figure}

The total number $N_C$ of carbon atoms in a cone structure may be estimated by dividing the cone surface area by half of the hexagonal cell's surface,
\begin{equation}
N_C = [4\pi/(3\sqrt{3})] (1-n_w/6) (r_D/a_{CC})^2 \,\,,
\end{equation}
where the disclination number $n_w$ corresponds to the integer number of $\pi/3$ wedge-sections eliminated from the disk structure, and $r_D$ is the cone generatrix [see Figure \ref{fig:pictorial-view}]. The nanocone disclination angle is given by $n_w \pi/3$. For example, for $n_w=1$ and $r_D=1\,\mu\text{m}$, the CNC has $\approx 10^8$ atoms. By extracting an integer number $n_w$ of $\pi/3$ sections from a carbon-disk [cf. Figure \ref{fig:pictorial-view}], it is possible to construct up to 5 different closed cones. For $n_w=1$, the cone angle is $2\theta_1=112.9^\circ$, corresponding to the flattest  possible cone. In this case, $h/r_C= 0.66$ and $h/r_D=0.55$.

In this work, finite-size systems (from two hundred up to five thousand atoms) are studied by performing direct diagonalizations of the stationary wave equation in the framework of a first-neighbor tight-binding approach. Each carbon atom has three nearest neighbours, except the border atoms for which dangling bonds are present. The overlap integral $s$ is considered different from zero. As we will show later, this has important effects on the cone energy spectrum.

Although some of the graphene electronic properties are present in the CNCs, deviations are always manifested as a consequence of the different atomic arrangements, the finite-size of the nanocones, and also the possible point symmetry of the distinct cones.
In the absence of external fields, the calculated density of states (DOS) shows a peak at the Fermi energy  and the local density of states (LDOS) shows that electron states are localized at the cone base.
On the other hand, the symmetries observed in the LDOS at different energies allow a systematic description of the electronic structure and selection rules of optical transitions driven by polarized radiation.
Unlike the nanodisk, the presence of topological disorders in nanocones involve a deviation from the electrical neutrality at the apex and at the zigzag edges.

\section{\label{sec:theo-FW}Theoretical Framework}

In what follows, we  present results for $n_w=\{0,1,2\}$, corresponding to CND and CNCs whose disclination angles are $60^\circ$ and $120^\circ$.
For those systems, the $sp^2$ hybridization may be neglected. The electronic wave function may be written as
\begin{equation}
| \Psi\rangle = \sum_{j=1}^{N_C} C_j | \pi_j\rangle
\, \,,
\label{eq:wavefunc}
\end{equation}
where the $|\pi_j\rangle$ denote the atomic orbitals $2p$ at site $\vec R_i$.
Note that the overlapping between neighbouring orbitals prevents the set consisting of the ket ${|\pi_j\rangle}$ to be an orthogonal basis.
Only in the ideal case of zero overlap $s=0$, the coefficients $C^0_j$ in $| \Psi\rangle = \sum_{j=1}^{N_C} C^0_j | \pi^0_j\rangle$ might be considered equal to the discrete amplitude probability $\langle \pi^0_j|\Psi\rangle$ to find an electron at the $j$-th atom (described by the one electron state $|\Psi\rangle$).
We use the $s\neq 0$ basis, $| \pi_j \rangle$, to construct the eigenvalue equation and the $| \pi^0_j \rangle$ base to calculate the properties related to discrete positions.
Of course, to relate both bases it is required to know the $\langle \pi^0_j | \pi_j \rangle$ projection.

We define a $N_C \times N_C$ matrix $\Delta^{(1)}$ relating the nearest neighbouring atomic sites $i,j$,
\begin{equation}
\label{eq:delta1}
\Delta^{(1)}_{ij} =
\left\{
\begin{matrix}
1, & (i,j) \text{ are n.n.} \\
0, & \text{otherwise}
\end{matrix}
\right\}\,.
\end{equation}
Similarly,
\begin{equation}
\label{eq:delta0}
\Delta^{(0)}_{ij} =
\left\{
\begin{matrix}
1, & (i=j)\\
0, & \text{otherwise}
\end{matrix}
\right\}\,.
\end{equation}

The $S$ overlap matrix elements are then given by
\begin{equation}
\label{eq:overlap}
S_{ij} =\langle \pi_i | \pi_j \rangle = \Delta^{(0)}_{ij} + s \Delta^{(1)}_{ij}\, .
\end{equation}
The hopping matrix elements of the tight-binding Hamiltonian $\hat{H}=\hat{H}^{(at)}+\hat{V}$ are
\begin{equation}
\label{eq:V_ij}
\langle \pi_i | \hat V | \pi_j \rangle = t \, \Delta^{(1)}_{ij}
\, ,
\end{equation}
where $t$ is the hopping energy parameter.
Assuming the eigenvalue equation $\hat H^{(at)} \, | \pi_j \rangle = \varepsilon_{2p} | \pi_j \rangle $, the atomic matrix elements are
\begin{equation}
\label{eq:H^at_ij}
\langle \pi_i | \hat H^{(at)} | \pi_j \rangle = \varepsilon_{2p} \left[ \Delta^{(0)}_{ij} + s \Delta^{(1)}_{ij} \right]
\, ,
\end{equation}
and
\begin{equation}
\langle \pi_i | \hat H |\Psi\rangle = \varepsilon \langle \pi_i |\Psi\rangle \quad \,\,,\,\, 1\leq i \leq N_C\,\,.
\end{equation}

The resulting equation system may be written as a generalized eigenvalue problem $H \vec C = \varepsilon S \vec C$, where the column vector $\vec C$ contains the coefficient $C_j$,
\begin{equation}
\label{eq:gen-eig-prob}
\left[ \varepsilon_{2p} (\Delta^{(0)} +s \Delta^{(1)}) +t\Delta^{(1)} \right] \vec C
= \varepsilon \left[ \Delta^{(0)} + s\Delta^{(1)} \right] \vec C.
\end{equation}
The general solution may be expressed in terms of the auxiliar variables $\vec C(0)$ and $\varepsilon(0)$, which satisfy
\begin{equation}
\label{eq:simple-eig-prob}
t \Delta^{(1)} \vec C(0) = \varepsilon(0) \vec C(0)\,.
\end{equation}
As $\vec C(0)$ also satisfies Eq. \eqref{eq:gen-eig-prob}, we obtain
\begin{equation}
\label{eq:E(s)}
\varepsilon = [\varepsilon_{2p} + (1+s\varepsilon_{2p}/t)\varepsilon(0)] / [ 1 + s \varepsilon(0)/t ]
\, .
\end{equation}
The orthogonality condition for the electronic states
\begin{equation}
\langle \Psi^{k} | \Psi^{l} \rangle  = \vec C^{k\dagger} \, S \, \vec C^{l}= \delta_{k,l}
\,
\end{equation}
implies that
\begin{equation}
\vec C = \frac{\vec C(0)}{\sqrt{{\vec C(0) }^\dagger \, S \, \vec C(0) }}
\, .
\end{equation}
For the calculation of the DOS we use a Lorentzian distribution
\begin{equation}
\label{eq:DOS}
\text{DOS}(\varepsilon) = 2\sum_{j=1}^{N_C} \delta(\varepsilon-\varepsilon^j) \approx 2\sum_{j=1}^{N_C}  \frac{\Gamma/\pi}{(\varepsilon-\varepsilon^j)^2+\Gamma^2}
\, .
\end{equation}

The LDOS is calculated in terms of the  discrete amplitude probability, $\langle  \pi^0_i | \Psi^{j} \rangle$,
\begin{equation}
\label{eq:LDOS}
\text{LDOS}(\vec R_i , \varepsilon) = 2 \sum_{j=1}^{N_C} |\langle \pi^0_i |\Psi^{j}\rangle |^2
\, \delta(\varepsilon - \varepsilon^j)\,\, ,
\end{equation}
where
\begin{equation}
\label{eq:wave-amplitude}
\langle  \pi^0_i | \Psi^{j} \rangle = \left[ \Delta^{(0)} + \frac{s}{2}\Delta^{(1)} \right] \vec C^{j}
\, ,
\end{equation}
as it is shown in Appendix A.

The local density of charge (LDOC) related to the $\pi$ electrons is calculated by assuming that the other 5 electrons and the 6 protons of the carbon atom act as a net charge $+e$.
Assuming zero temperature and the independent electron approximation, only the states $1 \leq j \leq n_F$ will be occupied, where
\begin{equation}
n_F=\left\{
\begin{matrix}
N_C/2 & , & N_C \text{ is even} \\
(N_C+1)/2 & , & N_C \text{ is odd} \, .\\
\end{matrix}
\right.
\end{equation}

Taking into account that the states below $n_F$ contribute with $-2e$ and the fact that the $n_F$ state contribution depends on the parity of the number of atoms in the system, the LDOC is written as
\begin{equation}
\label{eq:LDOC}
\text{LDOC}(\vec R_i) =\\
e \left[
1 - 2\sum_{j=1}^{n_F-1} |\langle \pi_i^0 | \Psi^j\rangle|^2
-\gamma|\langle \pi_i^0 | \Psi^{n_F}\rangle|^2
\right]
\end{equation}
with $\gamma$=0 and 1, for $N_C$ even and odd, respectively.

Optical absorption coefficients $\alpha_{\epsilon}(\omega)$ are calculated by considering perpendicular ($\hat {\epsilon_\bot}=\hat{\epsilon_x}, \hat{\epsilon_ y}$) and parallel ($\hat {\epsilon_{\|}}=\hat {\epsilon_z}$) polarizations, in relation to the cone axis,
\begin{equation}
\label{eq:coef:abs}
\alpha_{\hat{\epsilon}}( \omega) \propto
\frac{1}{\omega}
\sum_{i,j}
\left|
\langle \Psi^{i} | \, \hat{\epsilon} \cdot {\vec p} \, | \Psi^{j} \rangle
\right|^2
\delta\left[\varepsilon^j-\varepsilon^i -\hbar\omega \right]
\, \,,
\end{equation}
with ${\varepsilon^{i,j}}$ corresponding to the energies of occupied and unoccupied states, respectively.

The oscillator strength may be written in terms of the spatial operators ($\hat x$, $\hat y$ and $\hat z$) \cite{Pedersen2003}, i.e.,

\begin{equation}
\label{eq:<p_x>=<x>}
\langle \Psi^{i} | \hat p_z | \Psi^{j}  \rangle = \frac{m_e}{i \hbar} (\varepsilon^{j} - \varepsilon^{i} ) \langle \Psi^{i} | \hat z | \Psi^{j}  \rangle
\, \,\,,
\end{equation}
where $\langle \Psi^{i} | \hat z | \Psi^{j}  \rangle$ is calculated to first order in s,  using \eqref{eq:f_ij}  of Appendix A,
\begin{equation}
\label{eq:brackets:f(R)}
\langle \Psi^{i} | \hat{z} | \Psi^{j}  \rangle =
\vec C^{i\dagger} \left[ z + \frac{s}{2} \left( \Delta^{(1)} z + z \Delta^{(1)} \right) \right] \vec C^{j}
\, .
\end{equation}

\section{\label{sec:results}Results and Discussion}

\subsection{\label{sec:LDOS}Electronic Density of  States}

\begin{figure}
\includegraphics[width=\linewidth]{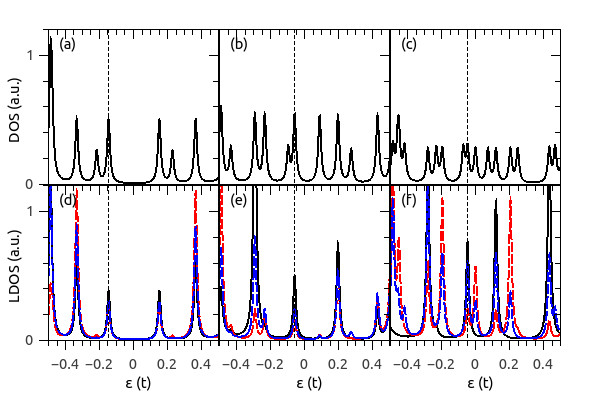}
\caption{\label{fig:DOS_LDOS_NC250_t/100}
(color online) DOS and LDOS for a $N_C=258$ nanodisk [(a) and (d)], a $N_C=245$ 1-pentagon nanocone [(b) and (e)], and a $N_C=246$ 2-pentagon nanocone [(c) and (f)].
LDOS curves (black-1, red-2, blue-3) match, for each system, the corresponding atoms shown in Figure \ref{fig:apice012pentagonos}.
Vertical lines in each panel indicate the position of the Fermi energy.}
\end{figure}

In what follows, we  present numerical results for systems composed of up to 5000 atoms.
In the limiting case of $N_C\to\infty$, the energy spectrum is in the range from  $\varepsilon_\text{min}=-3|t|/(1+3s)$ to $\varepsilon_\text{max}=+3|t|/(1-3s)$, the van Hove singularities occur at $\varepsilon_\text{vH}^v=-|t|/(1+s)$, $\varepsilon_\text{vH}^c = +|t|/(1 - s)$, and the Fermi energy is at $\varepsilon_\text{F}=0$. A $\Gamma=|t|/100$ broadening  and an overlap $s=0.13$ are assumed.
Different plots in Figure \ref{fig:DOS_LDOS_NC250_t/100} show the density of states averaged over the $N_C$ atoms and the LDOS for a CND [Figures \ref{fig:DOS_LDOS_NC250_t/100}(a) and \ref{fig:DOS_LDOS_NC250_t/100}(d)], a single-pentagon CNC [Figures \ref{fig:DOS_LDOS_NC250_t/100}(b) and \ref{fig:DOS_LDOS_NC250_t/100}(e)], and for a 2-pentagon CNC  [Figures \ref{fig:DOS_LDOS_NC250_t/100}(c) and \ref{fig:DOS_LDOS_NC250_t/100}(f)], for $N_C= 258, 245$, and  $246$, respectively.
All results are shown in an energy range around $\varepsilon_\text{2p}=0$.

\begin{figure}
\includegraphics[width=\linewidth]{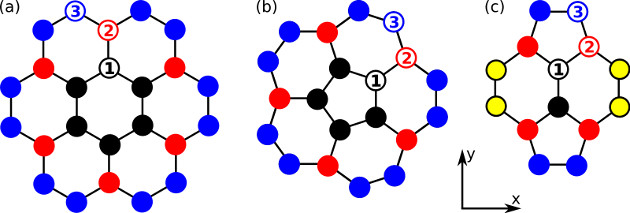}
\caption{
\label{fig:apice012pentagonos}
Pictorial view of (a) a nanodisk center, (b) a 1-pentagon nanocone apex, and (c) a 2-pentagon nanocone apex. Atoms with different colors-numbers indicate different point symmetries for each system.}
\end{figure}

\begin{figure}
\includegraphics[width=\linewidth]{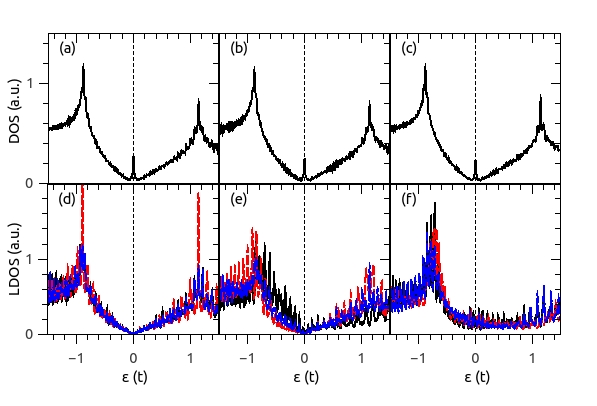}
\caption{\label{fig:DOS_LDOS_NC5000_t/100}
(color online) DOS and LDOS for a $N_C=5016$ nanodisk [(a) and (d)],  a $N_C=5005$ 1-pentagon nanocone [(b) and (e)], and a $N_C=5002$ 2-pentagon nanocone [(c) and (f)]. LDOS curves (black-1, red-2, blue-3) match, for each system, the corresponding atoms shown in Figure \ref{fig:apice012pentagonos}. Vertical lines in each panel indicate the position of the Fermi energy.}
\end{figure}

\begin{figure*}
\includegraphics[width=\linewidth]{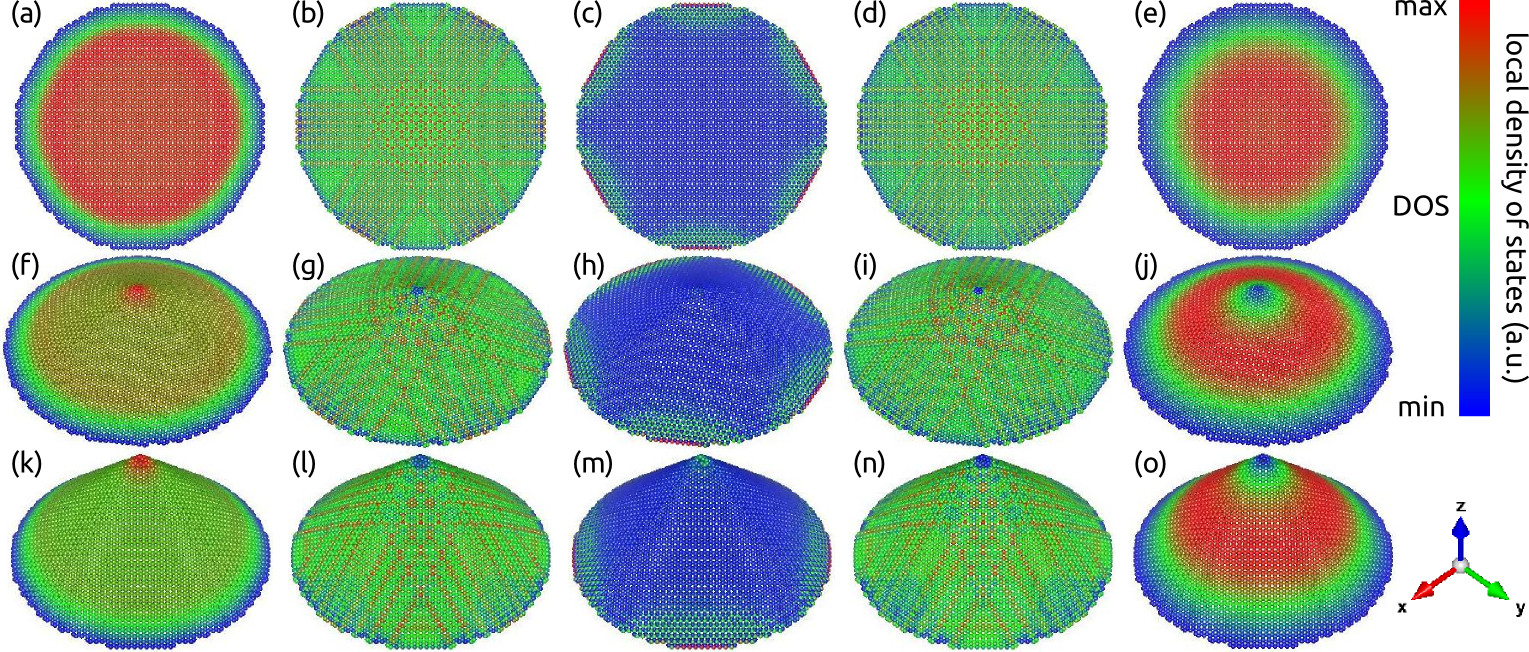}
\caption{\label{fig:LDOSdisk0}(Color online)
LDOS in arbitrary units for a 5016-atom nanodisk [panels (a) to (e)], a 5005-atom nanocone  with one pentagon at the apex [panels (f) to (j)], and a 5002-atom nanocone with 2 pentagons at apex [panels (k) to (o)]. The considered energies are (a,f,k) $\varepsilon_\text{min}$, (b,g,l) $\varepsilon^v_\text{vH}$, (c,h,m) $\varepsilon_\text{F}$, (d,i,n) $\varepsilon^c_\text{vH}$, and (e,j,o) $\varepsilon_\text{max}$. The LDOS is measured with respect to the mean LDOS which is equal to the DOS at the considered energy.}
\end{figure*}

As expected for small finite systems, the DOS, LDOS, and the position of the Fermi energy depend on the number of atoms considered in the numerical calculation and on their characteristic geometries \cite{Berber2000,Charlier2001,Navia2005}. A remarkable difference between CND and CNCs structures is the existence of a finite DOS above the Fermi level for nanocones. This clear metallic character of the DOS for nanocones is  more robust for the 2-pentagon CNC \cite{Charlier2001,Chao2009}. One may remark that this feature comes out from a symmetry break induced by the presence of topological defects in the CNC lattices, which generates new states above the Fermi energy, not  present in the CND structure. The contributions to the DOS coming from the apex atoms states are apparent in the LDOS of Figures \ref{fig:DOS_LDOS_NC250_t/100}(e) and \ref{fig:DOS_LDOS_NC250_t/100}(f). Also notice that for the 2-pentagon case, in which  there is a large topological disorder, the LDOS spectra exhibit significant  differences depending on the point symmetry of the considered atom (cf. Figure \ref{fig:apice012pentagonos}).

For increasing number of atoms, the total DOS  for the different nanostructures are very similar to the corresponding DOS of a graphene layer, except for the edges states which show up as a peak at the Fermi energy, as shown in Figures \ref{fig:DOS_LDOS_NC5000_t/100}(a), \ref{fig:DOS_LDOS_NC5000_t/100}(b), and \ref{fig:DOS_LDOS_NC5000_t/100}(c). It is interesting to note that the apex atomic states do not contribute to the total DOS near the Fermi energy  but mainly  near the graphene-like van Hove peaks. Notice that in the case of two-pentagon nanocones the LDOS at the tip exhibits a robust metallic character.

To analyse the finite-size effects and the role played by the different symmetries of the cone-tip sites, we depict LDOS contour plots for the three studied structures by considering some characteristic  energies: the minimum  energy, the  resonant peak below the  Fermi energy, the Fermi energy, the  resonant peak above  the Fermi energy, and the maximum energy. Figure \ref{fig:LDOSdisk0} illustrates the example of a CND with 5016 atoms (top row),  a single-pentagon CNC with 5005 atoms (middle row), and a 2-pentagon CNC with 5002 atoms (bottom row). The electronic states corresponding to energies at the band extrema have the largest wavelength compared to the characteristic size of the system. In this way, the details of the lattice become less important and the states exhibit azimuthal symmetry. An interesting feature for the nanocones is that at these energies the apex corresponds to a node for the maximum energy and an antinode for the minimum energy, respectively. On the other hand, the states at the Fermi energy are localized at the cone border, mainly at the zigzag edges as it is clearly shown in Figures \ref{fig:LDOSdisk0}(c), \ref{fig:LDOSdisk0}(h), and \ref{fig:LDOSdisk0}(m). For the states whose energy is near to the van Hove peaks, the LDOS reflect  the symmetries of each system, i.e., for CND the $2\pi/6$-rotation symmetry and $12$ specular planes [cf. Figures \ref{fig:LDOSdisk0}(b) and \ref{fig:LDOSdisk0}(d)],
for a single-pentagon CNC there is a $2\pi/5$-rotation symmetry and $5$ specular planes [cf. Figures \ref{fig:LDOSdisk0}(g) and \ref{fig:LDOSdisk0}(i)], and for a two-pentagon CNC, there is a $\pi/2$ rotation symmetry and $2$ specular planes [cf. Figures \ref{fig:LDOSdisk0}(l) and \ref{fig:LDOSdisk0}(i)].

\subsection{\label{sec:LDOC} Electric Charge Distribution}

The electric charge per site, in terms of the fundamental charge $e$, was obtained using Eq. \eqref{eq:LDOC}. Results for the electric charge distribution for CNDs indicate that all the atomic sites preserve the charge neutrality, i.e., LDOC=0.  For the CNCs, however, the atoms at the apex acquire negative charge and the atoms around the cone base exhibit positive charges at the zigzag edges. As $N_C$ increases, the LDOCs at the apices, for the two studied CNC structures, tend to the asymptotic values shown in table \ref{tab:ldoc},  which are in good agreement with the values reported by Green method calculations \cite{Tamura1994,Tamura1997}.

\begin{figure}
\includegraphics[width=\linewidth]{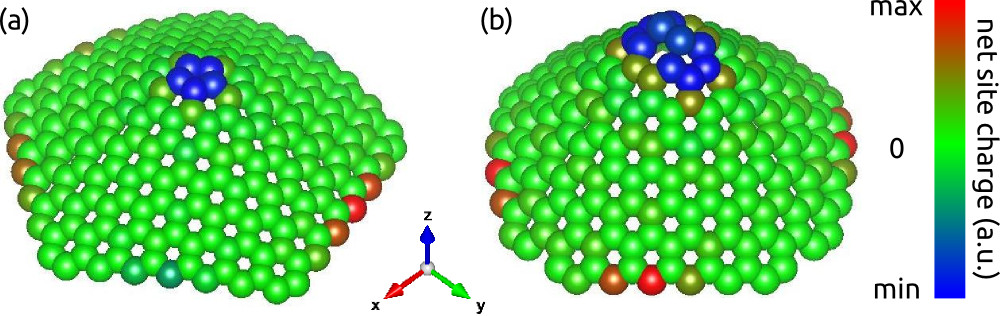}
\caption{\label{fig:ldoc_nw12_NC5000}(Color online)
Electric charge distribution in neutral CNCs, for a single-pentagon structure with 245 atoms (a) and two-pentagon system with 246 atoms (b).}
\end{figure}

Figure \ref{fig:ldoc_nw12_NC5000} depicts  the LDOC for the two types of CNC structures, showing that  the nonequilibrium of the charge distribution is restricted to the apex  and edge regions: electric neutrality is found at all the other surface sites. The values found for the LDOC at the apex regions are found to be independent of the size of the cones whereas this is not true for the edge states. When the number of atoms of the CNC structure is even, the edge-states LDOC exhibits the same symmetry of the cone. For odd $N_C$, the Fermi level is occupied by a single electron and then the LDOC at the edge states reflect the broken symmetry.
\begin{table}
\center
\begin{tabular}{|c|c|c|c|c|}
\hline
sites       &1    &2   &3   &max      \\
\hline
1-pentagon  &-0.071e   &+0.014e   &-0.059e   &+0.042e   \\
\hline
2-pentagon  &-0.055e   &-0.067e   &-0.066e   &+0.076e   \\
\hline
\end{tabular}
\caption{\label{tab:ldoc} LDOC (fundamental charge units) at some relevant atoms in the cone apices shown in Figure \ref{fig:apice012pentagonos}(b) and (c). The maximum (max) value occurs at the zigzag edge of each system.}
\end{table}

\subsection{\label{sec:absorption}Absorption Spectra}

\begin{figure}
\center
\includegraphics[width=7.1cm]{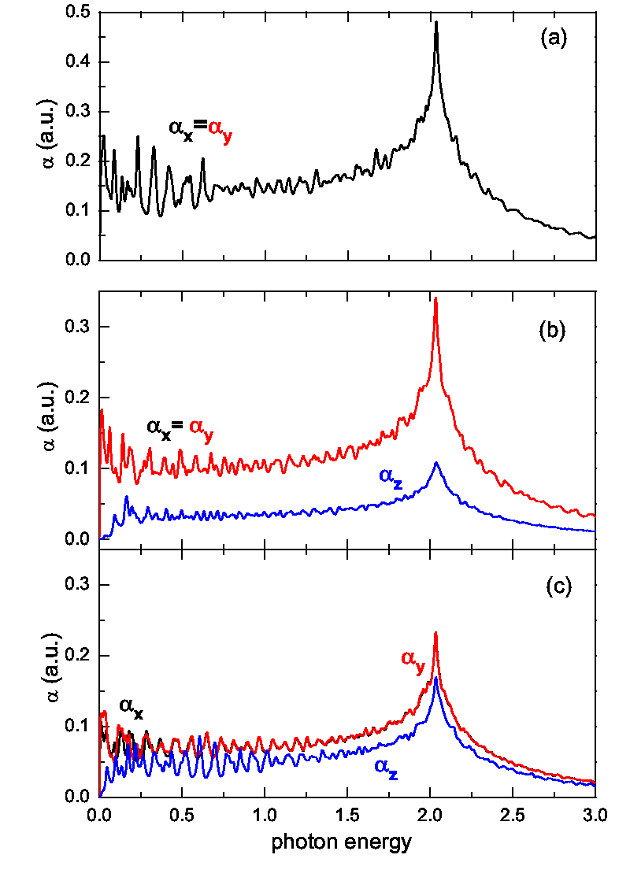}
\caption{\label{fig:abs5000}(Color online)
Absorption coefficient for $x$ (black curves), $y$ (red curves) and $z$ (blue lines) polarizations for (a) a nanodisk with 5016 atoms, (b) a single-pentagon nanocone composed of 5005 atoms, and (c) a two-pentagons nanocone with 5002 atoms. The photon energies are given in units of $\hbar \omega/t$.}
\end{figure}
We have also calculated the absorption coefficient for the CND and CNC structures, for different photon polarizations.
Figure \ref{fig:abs5000} shows the results for the absorption coefficients $\alpha_x$ and $\alpha_y$, for  polarization perpendicular to the cone axis, and $\alpha_z$ for  parallel polarization. Calculated results are shown for a nanodisk composed of 5016 atoms, a single-pentagon nanocone with 5005 atoms, and a 2-pentagon nanocone with 5002 atoms. For   the case of large CNDs, the  spectra present the general features observed for the absorption of a  graphene monolayer.
In the infrared region the absorption coefficient of a graphene monolayer is expected to be strictly constant \cite{Absorption1}, whereas for higher energies the spectrum shows a strong interband absorption peak coming from transitions near the M point of the Brillouin zone of graphene  \cite{Absorption2}. The main difference for a finite CND is a departure from a completely frequency-independent behavior for low energies, where the absorption coefficient  shows oscillations as a function of the photon energy instead of a constant value.  This is a consequence of the border states that are manifested as a peak in the total DOS at the Fermi energy \cite{Absorption3,Zhang2008}.  For CNCs, the general behaviour is the same as for  nanodisks,  except for the dependence   of  the absorption  on the photon  polarization, in particular for low energies. Furthermore, the main absorption peaks for different polarizations occur when the photon energy is equal to the energy  between the two DOS van Hove-like peaks (cf. Figure \ref{fig:DOS_LDOS_NC5000_t/100}). Notice that the overlap integral $s\ne 0$ leads to an energy shift of the main resonant absorption peak given by $\delta\approx 2 s^2 |t|/(1-s^2)\approx 100$ meV.  This is a significant value for actual experimental measurements.

Concerning the different polarization directions, one should notice that, as occurs in $C_{6v}$ symmetric systems,  $\alpha_z=0$ and $\alpha_x=\alpha_y$ for the nanodisk. On the other hand, the absorption coefficients for the different cones studied (single and two pentagons) are finite for parallel polarization,  and it depends on the structure details: as $\alpha_z$  increases for a two-pentagon CNC structure, $\alpha_{x,y}$ decreases.
Due to the lack of $\pi/2$- rotation symmetry, one should expect, in principle, different results for x- and y-polarizations for any nanocone. However, such difference is observable just for the absorption coefficient of the two-pentagon CNC system,  mainly in the range of low photon energies.
 The fact that $\alpha_x=\alpha_y$, for the case of 1-pentagon CNC structure, may be explained using similar symmetry arguments applied to $C_{6v}$ symmetry  dots\cite{Zhang2008}, extended to the $C_{5v}$ symmetric cones. In the case of a 2-pentagon CNC, the apex exibits a $C_{2v}$ symmetry, preventing the cone to be a $C_{4v}$ symmetric system. As the apex plays a minor role, $\alpha_x$ and  $\alpha_y$ will be slightly different.  
 A large difference between the $\alpha_z$ and the $\alpha_{x,y}$ CNC absorption spectra occurs in the limit of low radiation energy. The $\alpha_z$ coefficient goes to zero as $\hbar\omega\to0$ whereas $\alpha_{x,y}$ shows oscillatory features. The behaviour of the absorption for parallel polarization is due to the localization of the electronic states at the atomic sites around the cone border. As the spatial distribution of those states are restricted to a narrow extension along the z coordinate,  the $z$ degree of freedom is frozen for low excitation energies.
\begin{figure}
\center
\includegraphics[width=7.1cm]{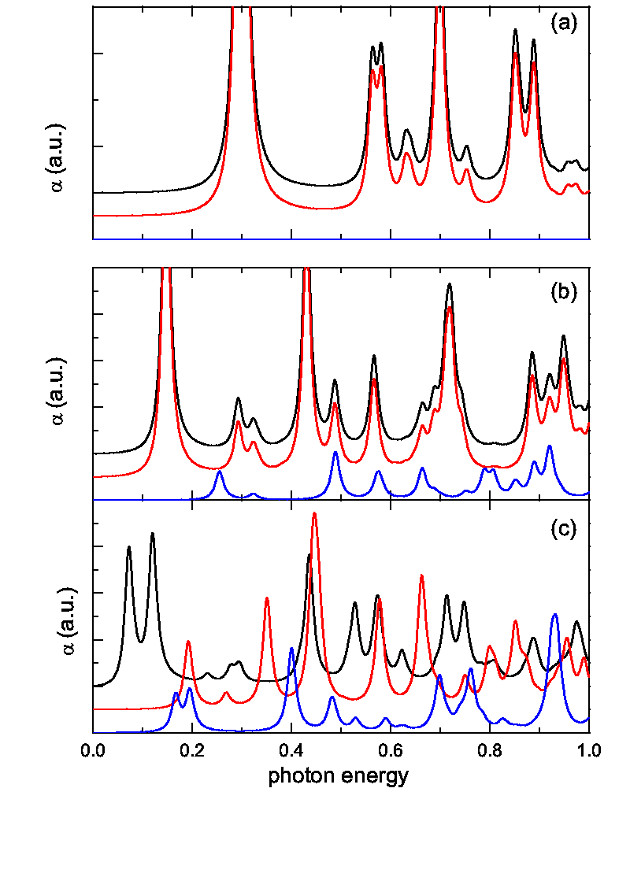}
\caption{\label{fig:abs0250}(Color online)
Absorption coefficient for $x$ (black curves), $y$ (red curves) and $z$ (blue lines) polarizations for (a) a nanodisk with 258 atoms, (b) a single-pentagon nanocone composed of 245 atoms, and (c) a two-pentagon nanocone with 246 atoms. The photon energies are given in units of $\hbar \omega/t$. Curves in each panel are vertically shifted, for better visualization of different polarization results.}
\end{figure}

The dependence of the absorption spectra  on the geometrical details of the different structures is more noticeable for  finite-size nanostructures. This   can be seen in Figure \ref{fig:abs0250} which depicts the absorption coefficients for the CND  composed of 258 atoms,  the single-pentagon CNC with 245 atoms, and the 2-pentagon CNC with 246 atoms. The degeneracy of the x- and y- polarization spectra is apparent for the smaller one-pentagon nanocone, as expected due to symmetry issues. On the other hand, the symmetry reduction for the 2-pentagon structure leads to  a rich absorption spectra, exhibiting peaks at different energies  and with  comparable weights for  distinct polarizations. In that sense, one may propose absorption experiments as an alternative route to distinguish between different nanocone geometries.

\section{\label{sec:conclusions}Conclusions}
 Here we have presented a theoretical study on the electronic properties of nanodisks and nanocones in the framework of a tight-binding approach.  We have proposed a discrete position approximation to describe the electronic states which  takes into account the effect of the overlap integral to first order.  While the $|\pi\rangle$ base keeps the phenomenology of the overlap between neighboring atomic orbitals, the  $|\pi^0\rangle$ base allows the construction of diagonal matrices of position-dependent operators.  A transformation rule was set up to take advantage of these two bases scenarios.

We have investigated the effects on the DOS and LDOS, of the size and topology of CND and CNC strutures.
We have found that both total and local density of states sensitively depend on the number of atoms and characteristic geometry of the structures.
One important aspect is the fact that cone and disk borders play a relevant role on the LDOS at the Fermi energy. For small finite systems the presence of states localized in the cone apices determines  the form of the DOS close to the Fermi energy. The observed features indicate that small nanocones could present good field-emission properties. This is corroborated by the calculation of the LDOC, that indicates the existence of finite charges at the apex region of the nanocones. For large  systems, the contribution to the DOS near the Fermi level is mainly due to  states localized in the edges  of the structures wheres for other energies, the DOS  exhibits similar features to the case of a graphene lattice.

The absorption coefficient  for different CNC structures are calculated  and we have found a peculiar dependence on  the photon  polarization in the infrared range for the investigated systems. The symmetry reduction of the 2-pentagon nanocones causes the formation of highly structured absorption spectra, with  comparable weights for  distinct polarizations. The breaking of the degeneracy for different polarizations is found to be more pronounced for small nanocones. Absorption experiments may be used as natural measurements to distinguish between different nanocone geometries.

\section*{Acknowledgements}
This work  was supported by Fondecyt grant 1100672 and USM internal grant 11.11.62. P. Ulloa thanks DGIP and Mecesup PhD scholarships and the warm hospitality of the Departamento de Fisica da Materia Condensada da Universidade Estadual de Campinas and Instituto de F\'isica da Universidade Federal Fluminense.
Special thanks to Professor Patricio Vargas for his helpful advices. 

\appendix
\section{}

A discrete position scheme in terms of the $| \pi^0_j\rangle$ states was used to represent functions of the position given in terms of the atomic base, since they satisfy the same properties of the position states, i.e., orthogonality
\begin{equation}
\label{eq:<Ri|Rj>}
\langle \pi^0_i | \pi^0_j \rangle = \Delta^{(0)}_{ij} \, ,
\end{equation}
and completeness
\begin{equation}
\label{eq:identity}
\sum_{k=1}^{N_C}  | \pi^0_k \rangle \, \langle \pi^0_k | = \hat{1} \,
\end{equation}
in a $N_C$-dimensional subspace. The identity operator may also be constructed using the $s\neq 0$ base as
\begin{equation}
\hat 1 = \sum_{k,l}|\pi_k\rangle (S^{-1})_{kl} \langle \pi_l |
\, ,
\end{equation}
with the $S^{-1}\approx \Delta^{(0)} - s \Delta^{(1)} + O(s^2)$ matrix being different from the $N_C \times N_C$ identity matrix $\Delta^{(0)}$.

 We take $|\pi^0\rangle$ as the discrete position state and assume that the matrix elements $f^{R}_{ij}$ of position-dependent functions $f (\vec R)$ are known in the $s=0$ representation,
\begin{equation}
f^{R}_{ij} = \langle \pi^0_i | \hat f | \pi^0_j \rangle = f (\vec R_j) \, \Delta^{(0)}_{ij}
\, .
\end{equation}
Differently from the $f^R$ matrices, $f$ matrices in the $s\neq 0$ representation
\begin{equation}
f_{ij}=\langle \pi_i | \hat f | \pi_j \rangle \,
\end{equation}
are not diagonal. However, by performing the similarity transformation
\begin{equation}
\label{eq:similarity}
\langle \pi_i | \hat f | \pi_j \rangle = \sum_{k,l} \langle \pi_i | \pi^0_k \rangle \, \langle \pi^0_k | \hat f | \pi^0_l \rangle \, \langle \pi^0_l | \pi_j \rangle
\, ,
\end{equation}
we may obtain the unknown $f$ matrix in terms of the known $f^{R}$ matrix, provided the transformation rule between the $\pi^0$ and $ \pi$ bases is known.
By assuming $\langle \pi^0_i | \pi_j \rangle = \bar{\alpha} \Delta^{(0)}_{ij} + \bar{\beta} \Delta^{(1)}_{ij}$, the $s\neq 0$ representation may be found. The coefficients $\bar\alpha$ and $\bar\beta$ are obtained by using the identity \eqref{eq:identity} into Eq. \eqref{eq:overlap},
\begin{equation}
\label{eq:Sij+O(s^2)}
\sum_{k=1}^{N_C} \langle \pi_i | \pi^0_k \rangle \, \langle \pi^0_k | \pi_j \rangle = \Delta^{(0)}_{ij} + s \Delta^{(1)}_{ij} + O(s^2) \, ,
\end{equation}
and, to first order in $s$, ($\bar\alpha=1$ and $\bar\beta=s/2$) we have
\begin{equation}
\label{eq:basis:change}
\langle \pi^0_i | \pi_j \rangle = \Delta^{(0)}_{ij} + (s/2) \, \Delta^{(1)}_{ij}
\, .
\end{equation}
By replacing \eqref{eq:basis:change} in \eqref{eq:similarity}, one obtains
\begin{equation}
\label{eq:f_ij}
f_{ij} = \left[ f^{(R)} + \frac{s}{2} \left( f^{(R)} \Delta^{(1)} + \Delta^{(1)} f^{(R)} \right)  \right]_{i,j} + O(s^2)\,\,\,
\end{equation}
as the matrix elements of a position-dependent function in the $\pi$-base.

\bibliographystyle{elsarticle-num}
\bibliography{cncs}

\end{document}